\documentclass[12pt]{article}
\usepackage{amssymb}
\usepackage{amsmath,amsfonts}
\usepackage{hyperref}
\usepackage{cancel}
\usepackage[normalem]{ulem}
\usepackage{color}
\usepackage{graphicx}
\usepackage{subfigure}
\usepackage{ragged2e}
\usepackage{tikz}
\usepackage{authblk}
\usepackage{wasysym}
\usepackage[bottom]{footmisc}

\numberwithin{equation}{section}

\definecolor{blue-violet}{rgb}{0.54, 0.17, 0.89}
\definecolor{PineGreen}{cmyk}{0.92, 0, 0.59, 0.25}
\definecolor{OliveGreen}{cmyk}{0.64, 0, 0.95, 0.40}
\definecolor{RawSienna}{cmyk}{0, 0.72, 1, 0.45}
\definecolor{Gray}{cmyk}{0, 0, 0, 0.50}
\definecolor{MidnightBlue}{cmyk}{0.98, 0.13, 0, 0.43}
\definecolor{Orange}{cmyk}{0, 0.61, 0.87, 0}
\definecolor{LimeGreen}{cmyk}{0.50, 0, 1, 0}
\definecolor{Green}{cmyk}{1, 0, 1, 0}

\begin{document}
\title{\vspace{-30pt} {\sc Celestial closed strings at one-loop}\vspace{10pt}}
\author[a]{\large{Anthonny F. Canazas Garay}}
\author[b]{\large{Gaston Giribet}}
\author[c]{\large{Yoel Parra-Cisterna}}
\author[d]{\large{Francisco Rojas}}

\vspace{30pt}

\affil[a]{{\small\textit{Departamento de Cs. F\'isicas, Facultad de Cs. Exactas, Universidad Andres Bello, \newline Sazi\'e 2212, Santiago, Chile}}}
\affil[b]{{\small\textit{Department of Physics, New York University. 726 Broadway, New York,
NY10003, USA}}}
\affil[c]{{\small\textit{Instituto de F\'isica, Pontificia Universidad Cat\'olica de Chile, Av. Vicu\~na Mackenna 4860, Santiago, Chile}}}
\affil[d]{{\small\textit{Facultad de Ingenier\'ia y Ciencias, Universidad Adolfo Ib\'a\~nez, Santiago, Chile}}}

\date{}

\maketitle

\vspace{1cm}
\thispagestyle{empty}

\begin{abstract}
In this paper we continue our investigation of superstring scattering amplitudes in the conformal basis. We focus on the case of four graviton scattering processes at 1-loop in \emph{closed} superstring theory. We write the expression for such a process in the celestial variables and confirm previous expectations. In particular, we find the adequate overall factorization of the $\alpha'$ dependence which organizes the loop expansion of closed string celestial amplitudes. We also show that, at 1-loop, the field theory limit, when properly defined, commutes with the Mellin transform of the amplitudes for all values of the conformally invariant cross-ratio, something that had already been observed for gluon processes at 1-loop in open string theory and is to be compared with the tree-level computations. This indicates that many of the features satisfied for open string gluon amplitudes at tree and 1-loop levels are also universal properties of graviton celestial amplitudes in closed string theory. As a by-product, we also compute field theory graviton amplitudes at 1-loop in the conformal basis.
\end{abstract}

\newpage
\tableofcontents

\justify


\section{Introduction}

Celestial amplitudes describe scattering processes in the basis of boost eigenstates; they relate to the standard amplitudes in momentum space through a Mellin transform over the full range of energies of each of the external states participating in the collision \cite{Pasterski:2016qvg, Pasterski:2017kqt, Pasterski:2017ylz}. This makes celestial amplitudes to be sensitive to both ultraviolet and infrared effects. 

String amplitudes reproduce field theory amplitudes by taking the low energy limit of the former. Therefore, because of this UV/IR mixing, a natural question arises as to what is the appropriate field theory limit of string scattering amplitudes in the celestial basis. As pointed out by Stieberger and Taylor \cite{Stieberger:2018edy}, it is not possible to obtain such a limit by simply taking $\alpha' \to 0$. This question is equivalent to that of what is the interplay between the parameters that control the moduli space of the worldsheet, the kinematics, and the $\alpha' \to 0$ limit of the amplitude.\footnote{For other pioneering work on the relation between string amplitudes and the celestial sphere, see \cite{Stieberger:2013nha,Stieberger:2013hza}.} 

This question was first addressed in \cite{Stieberger:2018edy}, where they studied 4-point scattering amplitudes involving gluons in the type I superstring theory at tree-level in the celestial basis. There, the forward scattering limit of the stringy factor was identified as a limit in which celestial Yang-Mills 4-point function is recovered. The case of tree-level graviton amplitudes in the heterotic theory was also investigated in \cite{Stieberger:2018edy}. The analysis was extended in \cite{Castiblanco:2024hnq} to the case of 5-gluons at tree level, for which the string moduli space allows for more types of limits in the kinematic space. The celestial string amplitudes at 1-loop were investigated in \cite{Donnay:2023kvm}, where the case of gluons in the type I theory was analyzed in detail; see also \cite{Bockisch:2024bia}. It was shown in \cite{Donnay:2023kvm} that the Mellin transforms commute with the adequate limit in the worldsheet moduli space and reproduces the celestial 1-loop field theory gluon amplitude expressed in the worldline formalism\footnote{The fact that the 1-loop field theory amplitude appears in the worldline formalism is a natural feature since this formalism was originally inspired using string theory techniques.}. One of the characteristic features observed in all these studies of string amplitudes in the celestial basis is that, as a product of the Mellin transform when translating into the celestial basis, the dependence of $\alpha'$ in the amplitudes reduces to a simple overall factor whose exponent depends on the loop order and the number of uncompactified dimensions. Another interesting aspect of string amplitudes in celestial basis is that, in contrast to tree-level, at 1-loop the field theory limit is recovered for a wider range of the Mandelstam variables ratios, and not necessarily in the forward scattering limit. 
In order to investigate how general these features are, and to advance in understanding the reasons for the general properties of celestial amplitudes, in this work we will explore the case of graviton amplitudes at 1-loop in closed superstring theory. Given the inherent difficulty of the field theory description of gravity, this will allow us to understand which features of those observed in the Yang-Mills case also appear in this theory and which are exclusive to gauge theory. We will prove that several of the features observed in \cite{Stieberger:2018edy, Castiblanco:2024hnq, Donnay:2023kvm} are also found in the 1-loop graviton case, even though the limit works differently. This permits us to have a larger picture of the problem. We will show explicitly that the Mellin transform leading to the celestial basis commutes with the field theory limit if the latter is appropriately taken. We will also show that the $\alpha'$ dependence factors out as it does in the cases investigated before, showing that it is a generic fature of string amplitudes in the conformal basis.

The paper is organized as follows: In section 2, we will review the graviton amplitudes in quantum field theory at tree-level and 1-loop. In section 3, we will revise the analogous computation in closed superstring theory, collecting results and presenting the expressions in a way that will result useful for our analysis. Celestial string amplitudes will be discussed in section 4, where  we will explicitly compute the celestial amplitude for four gravitons at 1-loop, and explore its field theory limit. We will compare it with the result obtained by calculating the celestial amplitude coming from supergravity $\mathcal{N}=8$. Section 5 will contain our final remarks and conclusions.

\section{Graviton amplitudes}

Scattering amplitudes are usually expressed in terms of the Mandelstam variables, $ s_{ij}$, constructed from Lorentz invariants from the momenta of the $n$ external particles. Here, $k_{i}$ will denote the momentum of the $i^{\text{th}}$ external particle with components $k_{i}^{\mu}$ ($\mu=0,1,2,...,D-1$) in $D$ non-compact space-time dimensions; here $D=4$. For massless particles, $(k_{i})^{2}=0$, the Mandelstam variables are
\begin{equation}\label{Mandelstamstu2}
    s_{ij}\equiv-(k_{i}+ k_{j})^{2}=-2 k_{i}\cdot k_{j}\,,
\end{equation}
where, in the 4-point case, $s_{12}\equiv s, s_{23}\equiv t$, and $s_{13}\equiv u$, with $s+t+u=0$. The 4-point amplitude takes the general form
\begin{eqnarray}
    \mathcal{A}\left(\left\{k_{i}, \ell_{i}\right\}\right)=A\left(s, t, u, \left\{\ell_{i}\right\}\right) \delta^{(4)}\left(\sum_{i=1}^{4}k_{i}\right) \,,
\end{eqnarray}
where the $\{ \ell_{i}\}$ correspond to the helicities of the external particles with, for instance $\pm 1$ for gluons, $\pm 2$ for gravitons. The delta function, corresponding to total momentum conservation, has also been explicitly written.  

The Mandelstam variables \eqref{Mandelstamstu2} encode relevant kinematical information of the scattering process. The Lorentz invariant $s$ corresponds to the square of the total energy of the collision when computed in the center-of-mass frame. The region $s>0$ is thus usually referred to as the {\it physical region}.\footnote{In this work, this distinction becomes important because string amplitudes must usually be defined in the {\it non-physical region} in order to have well-defined integral representations. After these amplitudes are computed, one analytically continues the results at the very end. See, \emph{e.g.}, the discussion in Section 6.4 of \cite{Polchinski:1998rq}.} The scattering angle $\theta$ in the center-of-mass frame is obtained from the relation
\begin{eqnarray}\label{cross-ratio 1}
    r\equiv-\frac{s}{t}=\csc ^{2}\left(\frac{\theta}{2}\right)\,,
\end{eqnarray}
from where it follows that, in the physical region, $t<0$ and $r>1$. When written in terms of the usual parametrization for the external momenta in celestial variables (see equation \eqref{celestial_k}), $r$ becomes an invariant cross-ratio under global $SL(2,\mathbb{C})$ transformations on the cesphere.

In general relativity (GR), the amplitude for 4 gravitons is 1-loop renormalizable \cite{tHooft:1974toh}. Schematically, we can write
\begin{equation}
    A_{GR}(s, t, u)=A^{\textrm{tree }}_{GR}(s, t, u)+A^{1-\textrm{loop }}_{GR}(s, t, u)+\dots\,,
\end{equation}
where the ellipsis stand for higher-loop corrections. The general form of the tree-level contribution is
\begin{equation}\label{GR gravitons K}
    A^{\textrm{tree }}_{GR}(s, t, u)=\kappa_{4}^{2}\, \frac{K(\{k_{i}, \zeta^{\mu \nu}_{i}\})}{stu}\,
\end{equation}
with the 4-dimensional Planck length $\kappa_{4}=\sqrt{32\pi G}$ being the gravitational coupling constant. The kinematic factor $K (\{k_{i}, \zeta^{\mu \nu}_{i}\})$ is constructed from several contractions among the momenta $\{k_{i}\}$ and polarizations of the external gravitons, $\zeta^{\mu \nu}_{i}=\zeta^{\mu}_{i}\zeta^{\nu}_{i}$. Explicitly,
\begin{equation}
\begin{aligned}
&K (\{k_{i}, \zeta^{\mu \nu}_{i}\}) =\zeta_{1}^{\mu_{1} \mu_{2}}\zeta_{2}^{\nu_{1} \nu_{2}}\zeta_{3}^{\sigma_{1} \sigma_{2}}\zeta_{4}^{\lambda_{1} \lambda_{2}}K_{\mu_{1} \nu_{1} \sigma_{1} \lambda_{1}}K_{\mu_{2} \nu_{2} \sigma_{2} \lambda_{2}}\\
& \ \ \ \ \  \ =-\frac{1}{4}\left(s t \zeta_1 \cdot \zeta_3 \zeta_2 \cdot \zeta_4+s u \zeta_2 \cdot \zeta_3 \zeta_1 \cdot \zeta_4+t u \zeta_1 \cdot \zeta_2 \zeta_3 \cdot \zeta_4\right) \\
&\ \ \ \ \  \  +\frac{1}{2} s\left(\zeta_1 \cdot k_4 \zeta_3 \cdot k_2 \zeta_2 \cdot \zeta_4+\zeta_2 \cdot k_3 \zeta_4 \cdot k_1 \zeta_1 \cdot \zeta_3+\zeta_1 \cdot k_3 \zeta_4 \cdot k_2 \zeta_2 \cdot \zeta_3+\zeta_2 \cdot k_4 \zeta_3 \cdot k_1 \zeta_1 \cdot \zeta_4\right) \\
&\ \ \ \ \ \   +\frac{1}{2} t\left(\zeta_2 \cdot k_1 \zeta_4 \cdot k_3 \zeta_3 \cdot \zeta_1+\zeta_3 \cdot k_4 \zeta_1 \cdot k_2 \zeta_2 \cdot \zeta_4+\zeta_2 \cdot k_4 \zeta_1 \cdot k_3 \zeta_3 \cdot \zeta_4+\zeta_3 \cdot k_1 \zeta_4 \cdot k_2 \zeta_2 \cdot \zeta_1\right) \\
& \ \ \ \ \ \ +\frac{1}{2} u\left(\zeta_1 \cdot k_2 \zeta_4 \cdot k_3 \zeta_3 \cdot \zeta_2+\zeta_3 \cdot k_4 \zeta_2 \cdot k_1 \zeta_1 \cdot \zeta_4+\zeta_1 \cdot k_4 \zeta_2 \cdot k_3 \zeta_3 \cdot \zeta_4+\zeta_3 \cdot k_2 \zeta_4 \cdot k_1 \zeta_1 \cdot \zeta_2\right)\,.
\end{aligned}
\end{equation}
For a specific parameterization of momenta and polarizations, see for instance section V of reference \cite{Sannan:1986tz}. As in the case of Yang-Mills, one can also express the graviton amplitude in the spinor-helicity formalism. Here, we will focus on maximally helicity violating (MHV) amplitudes. At tree-level, using also spinor-helicity variables, such amplitudes take the form
 \cite{Dunbar:1994bn}
\begin{equation}\label{tree-level 4-gravitonGR}
A^{\textrm{tree }}_{GR}(-, -, +, +)=\frac{\kappa^{2}_{4}}{4}\left(\frac{\langle 12\rangle^4}{\langle 12\rangle\langle 23\rangle\langle 34\rangle\langle 41\rangle}\right)^2 \times \frac{s t}{u}\,.
\end{equation}
The contribution at 1-loop, expressed in terms of the Mandelstam variables, takes the (lengthier) form
\begin{equation}\label{1-loop 4-gravitonGR}
\begin{aligned}
\hspace{-0.5cm}A^{\textrm{1-loop}}_{GR}(-, -, +, +)=&\frac{i s t u\, \kappa_{4}^{2}(4 \pi)^\epsilon r_{\Gamma}}{4(4 \pi)^2} A^{\textrm{tree }}_{GR}\left(s, t, u\right) \left[\frac{2}{\epsilon}\left(\frac{\log (-u)}{s t}+\frac{\log (-t)}{s u}+\frac{\log (-s)}{t u}\right)\right. \\
& +\frac{2 \log (-u) \log (-s)}{s u}+\frac{2 \log (-t) \log (-u)}{t u}+\frac{2 \log (-t) \log (-s)}{t s} \\
& +\frac{(t+2 u)(2 t+u)\left(2 t^4+2 t^3 u-t^2 u^2+2 t u^3+2 u^4\right)\left(\log ^2(t /u)+\pi^2\right)}{s^8} \\
& +\frac{(t-u)\left(341 t^4+1609 t^3 u+2566 t^2 u^2+1609 t u^3+341 u^4\right) \log (t /u)}{30 s^7} \\
& \left.+\frac{1922 t^4+9143 t^3 u+14622 t^2 u^2+9143 t u^3+1922 u^4}{180 s^6}\right]\,,
\end{aligned}
\end{equation}
where $D=4-2\epsilon $ and 
\begin{equation}
r_{\Gamma}=\frac{\Gamma^2(1-\epsilon) \Gamma(1+\epsilon)}{\Gamma(1-2 \epsilon)}\,.
\end{equation}
The order $\mathcal{O}(\epsilon^{-1})$ term gives the usual gravitational divergence in $D= 4$. When computing a physical observable, such as the cross section, the infrared 1-loop divergence will eventually be canceled against soft radiation \cite{Donoghue:1999qh}, yielding an infrared-safe observable involving the four gravitons up to 1-loop in the perturbative expansion.

Since here we will be concerned with the field theory limit of type II superstrings compactified to $D=4$, it is instructive to write the expression for the 1-loop (MHV) amplitude in the case of $\mathcal{N}=8$ supergravity; namely \cite{Dunbar:1994bn}
\begin{equation}\label{1-loop 4-gravitonN=8}
\begin{aligned}
 A^{\text{1-loop}}_{\mathcal{N}=8}(-, -, +, +)&=i \kappa^{2}_{4} \frac{r_{\Gamma}(4 \pi)^\epsilon}{(4 \pi)^2} A^{\text {tree }}_{GR}\left(s, t, u\right) \, \left(\frac{2 s \log (-s)+2 t \log (-t)+2 u \log (-u)}{\epsilon}+\right.\\
&   \left. 2 s \log (-t) \log (-u)+2 t \log (-u) \log (-s)+2 u \log (-s) \log (-t)\frac{}{}\right)\,.
\end{aligned}
\end{equation}
This also admits to be written in terms of a sum of terms involving the hypergeometric function tending to a confluent point when $\epsilon \to 0$, namely, \scalebox{0.85}{$s_{ij}\,_2 F_1\left(-\epsilon,-\epsilon ; 1-\epsilon ; 1+(s_{ik}/s_{jk})^{\pm 1}\right)$}. Note  that, at order $\mathcal{O}(1/\epsilon )$, the sign in the argument of the logarithms of the Mandelstam variables can be inverted by using the condition $s+t+u+=0$.

\section{String amplitudes}

\subsection{Graviton amplitudes}

In this section we will follow the celestial holography program for the case of type II superstring processes for massless states in $D=4$ uncompactified dimensions. The simplest compactification scenario is to consider toroidal compactifications to 4 dimensions with radii $R=\sqrt{\alpha'}/a$, with $a$ being a dimensionless parameter. In the simplest case, when compactifying to $D$ dimensions, one could simply demand to set the remaining $10-D$ components of the external 10-momenta to zero, namely $k_{j}^{\mu}=(k_{j}^{0}, k_{j}^{1}, ... , k_{j}^{D-1}, 0,..., 0 )$ with $\mu = 0,1,2,...,9$. At the loop level, the strings still probe the extra dimensions and more care is needed when computing the integrations over the loop momenta. However, for the simple case of toroidal compatifications (due to the fact that there is no internal curvature), the effect is a simple modification of the integrand in the 1-loop string factor, which should now include a function $F_{2}(a, \tau)$ of $a$ and the modular parameter $\tau $; see \eqref{F2(a,tau)} below.\footnote{Note also that, since \eqref{F2(a,tau)} has no dependence on the Mandelstam variables, the compactification will not have a significant effect when going to the conformal basis.}

The scattering amplitude of four closed string gravitons, in the type II theory\footnote{The distinction between the chiralities of the type II theories is only relevant in the critical dimension, thus, in the toroidal compactification to $D=4$ we are using here, it is not necessary to distinguish between type IIA and type IIB amplitudes \cite{Green:2012pqa}.}, has the following form \cite{ Green:1982sw, Schwarz:1982, Green:2012pqa}
\begin{eqnarray}\label{massless amplitude type II}
    A_{II}(s, t, u)\hspace{-0.1cm}&=& A^{\textrm{tree}}_{II}+A_{II}^{1-\textrm{loop}}+\cdots\notag\,\\
    &=&\kappa_{10}^{2}\,K(\{k_{i}, \zeta^{\mu \nu}_{i}\})\left(g^{\textrm{tree}}_{II}(s, t, u)+g^{1-\textrm{loop}}_{II}(s, t, u)+\cdots\right)\,,
\end{eqnarray}
where $\kappa_{10}^{2}$ is the $10-$dimensional gravitational coupling constant. In this expression, the factor $g^{\textrm{tree}}_{II}$ represents the tree-level contribution of the strings, coming from integrating the graviton vertex operators on the Riemann sphere, modulo the stabilization of the conformal Killing group. This takes the Virasoro-Shapiro form
\begin{equation}\label{tree level}
    g^{\textrm{tree}}_{II}(s, t, u)=\frac{1}{stu}\frac{\Gamma(1-\alpha' s/4)\Gamma(1-\alpha' t/4)\Gamma(1-\alpha' u/4)}{\Gamma(1+\alpha' s/4)\Gamma(1+\alpha' t/4)\Gamma(1+\alpha' u/4)}\,.
\end{equation}
The 1-loop contribution, $g^{1-\textrm{loop}} _{II}$, results from integrating on the genus-1 surface; namely
\begin{equation}\label{one loop}
    g^{1-\textrm{loop}} _{II}(s, t, u)=\frac{\kappa^{2}_{10}}{\alpha'}\int_{\mathcal{F}}\frac{d\tau d\bar{\tau}}{\left(\textrm{Im}(\tau)\right)^{2}}[F_{2}(a, \tau)]^{10-D}F_{II}(\tau, s, t, u)\,,
\end{equation}
where $\mathcal{F}$ is the fundamental domain over the moduli space of the torus,
\begin{equation}\label{FundamentalDom. torus}
    \mathcal{F}=\left\{ -{1}\leq 2\textrm{Re}(\tau)\leq{1}, \hspace{0.5cm}\textrm{Im}(\tau)\geq0, \hspace{0.5cm}|\tau|^{2}\geq1\right\}\,.
\end{equation}
As anticipated, the function $F_{2}(a, \tau)$ appears as a consequence of the toroidal compactification of $10-D$ dimensions. This function is
\begin{equation}\label{F2(a,tau)}
    F_{2}(a, \tau)=a(\textrm{Im}(\tau))^{1/2}\sum_{m, n}\exp\left(-2\pi i m n \textrm{Re}(\tau)-\pi(a^{2}m^{2}+n^{2}/a^{2})\textrm{Im}(\tau) \right)\,.
\end{equation}
Function $F_{II}(\tau, s, t, u)$, on the other hand, is $SL(2, \mathbb{Z})$ invariant on the $\tau $-plane, given by the integral over the punctured-torus with marked points at $\nu_{i}$ in the domain
\begin{eqnarray}\label{nu's dominio}
    0\leq \textrm{Im}\nu_{i}\leq \textrm{Im}\tau\,,\hspace{1cm}-{1}\leq 2\textrm{Re}\nu_{i}\leq {1}\,.
\end{eqnarray}
Its explicit form is
\begin{eqnarray}\label{FII}
    F_{II}(\tau, s, t, u)&=&\left(\textrm{Im}(\tau)\right)^{-3}\int\prod^{3}_{l}d^{2}\nu_{l}\, \prod_{i<j}^{4}\left(\chi_{i j}\right)^{\alpha'k_{i}\cdot k_{j}}\,\notag\\
    &=&\left(\textrm{Im}(\tau)\right)^{-3}\int\prod^{3}_{l}d^{2}\nu_{l}\, \left(\frac{\chi_{1 2}\chi_{3 4}}{\chi_{1 3}\chi_{2 4}}\right)^{-\frac{\alpha'}{2}s}\left(\frac{\chi_{2 3}\chi_{1 4}}{\chi_{1 3}\chi_{2 4}}\right)^{-\frac{\alpha'}{2}t}\,,
\end{eqnarray}
where we have used the definition of Mandelstam variables \eqref{Mandelstamstu2} in the second line, and we have also used the short notation $\chi_{ij}\equiv \chi(\nu_{ij}, \tau)$ for the function
\begin{equation}\label{chifunction}
    \chi(\nu_{ij}, \tau)=2\pi \exp\left( -\frac{\pi(\textrm{Im}(\nu_{ij}))^{2}}{\textrm{Im}(\tau)}\right)\left|\frac{\vartheta_{1}(\nu_{ji}|\tau)}{\vartheta'_{1}(0|\tau)}\right|\,.
\end{equation}
The latter obeys the periodicity properties $\chi(\nu+1, \tau)=\chi(\nu+\tau, \tau)=\chi(\nu, \tau)$, compatible with the topology of the worldsheet. $\vartheta_{1}(\nu_{ji}|\tau)$ is the Jacobi theta function, which obeys
\begin{eqnarray}\label{psifunction}
   \frac{\vartheta_{1}(\nu_{ji}|\tau)}{\vartheta'_{1}(0|\tau)}
&=&\frac{\sin(\pi\nu_{ji})}{\pi}\prod_{n=1}^{\infty}\frac{\left(1-2q^{n}\cos(2\pi\nu_{ji})+q^{2n}\right)}{ (1-q^{n})^{2}}\,.
\end{eqnarray}
where $q=e^{2\pi i \tau}$. $\vartheta'_{1}(0|\tau)$ stands for the derivative of the Jacobi function with respect to the coordinates $\nu_{ji}=\nu_{j}-\nu_{i}$ evaluated at zero.

To make contact with the classical field theory result, one needs to express the tree-level contribution in \eqref{massless amplitude type II} in a form that turns out to be suitable to compare with \eqref{tree-level 4-gravitonGR}. In the limit $\alpha'\to 0$, considering \eqref{GR gravitons K} in $D=10$, one directly obtains \cite{Sannan:1986tz}
\begin{eqnarray}\label{tree level field theory}
     \lim_{\alpha'\to 0}A^{\textrm{tree}}_{II}(s, t, u)=\kappa_{10}^{2}\frac{K(\{k_{i}, \zeta^{\mu \nu}_{i}\})}{stu}=A^{\textrm{tree }}_{GR}(s, t, u)\,.
\end{eqnarray}
This allows to express the string amplitude \eqref{massless amplitude type II} conveniently as follows
\begin{eqnarray}\label{massless amplitude type II mejorada}
    A_{II}(s, t, u)\hspace{-0.1cm}=stu \,A^{\textrm{tree }}_{GR}(s, t, u)\left(g^{\textrm{tree}}_{II}(s, t, u)+g^{1-\textrm{loop}}_{II}(s, t, u)+\cdots\right)\,.
\end{eqnarray}

\subsection{The field theory limit}\label{Field limits of type II superstring}

In the momentum basis, the field theory limit is directly obtained by taking $\alpha'\to 0$ in the superstring amplitudes provided one performs the integration over the moduli space in the proper way. At tree-level, this is quite simple and yields \eqref{tree level field theory}. At 1-loop, this requires a little more care. Expression \eqref{one loop} involves the integration on the modular parameter of the genus-1 surface, $\tau$. Besides, it is accompanied by the factor \eqref{F2(a,tau)} accounting for the compactified directions. The right field theory limit thus corresponds to taking $\alpha'\to 0$ and the compactification radius $R\to 0$ while holding $\kappa_{D}$ and $a$ fixed.\footnote{Here we have chosen the same compactification radius for each extra dimension. While it is also possible to consider distinct radii for each compact dimension, such variations do not significantly impact the results of our computations.} This is controlled by the region $\textrm{Im}\tau\to \infty$ . In that limit, the result for the type II superstring theory with $D=4$ leads to that of the $\mathcal{N}=8$ supergravity, cf. \cite{ Green:1982sw, Schwarz:1982}. Explicitly, from \eqref{massless amplitude type II} we get
\begin{eqnarray}\label{fieldlimit final1}
     \lim_{\alpha'\to 0}g^{1-\textrm{loop}}_{II}(s, t, u)=\kappa_{D}^{2}\pi^{D/2-4}\Gamma(-\gamma)\left(\int _{0}^{1}[d\rho]\,[s\rho_{1}\rho_{2}+t\rho_{2}\rho_{3}+u\rho_{3}\rho_{1}+C]^{\gamma}+\textrm{perm}\right)\notag\,\\
\end{eqnarray}
where one identifies $\kappa^{2}_{D}=\kappa^{2}_{10}R^{D-10}$. The parameter $\gamma=-4+D/2$ is defined to have control on the limit of interest, $D\to 4$; it relates to the dimensional regularization parameter through $\epsilon=\gamma+2$. In (\ref{fieldlimit final1}), $C=t(\rho_{1}-\rho_{2})$ and ``perm'' stands for permutations of the $\rho_i$, considering all the order combinations. For example, for the order $0\leq \rho_{1}< \rho_{2}< \rho_{3}\leq \rho_{4}\equiv 1$, the integral is defined according to
\begin{equation}\label{integrals rho}
    \int_{0}^{1} [d\rho]=\int _{0}^{1}d\rho_{3}\int_{0}^{\rho_{3}}d\rho_{2}\int_{0}^{\rho_{2}}d\rho_{1}\,,
\end{equation}
with $C=t(\rho_{1}-\rho_{2})$. In general, for all possible permutations, one must consider
\begin{equation}\label{Aorderingrho}
    C=\left\{
                \begin{array}{ll}
                  t(\rho_{1}-\rho_{2})\,,\hspace{1cm}\rho_{1}<\rho_{2}<\rho_{3}\\
                  t(\rho_{1}-\rho_{3})\,,\hspace{1cm}\rho_{1}<\rho_{3}<\rho_{2}\\
                u(\rho_{2}-\rho_{1})\,,\hspace{1cm}\rho_{2}<\rho_{1}<\rho_{3}\\
                u(\rho_{2}-\rho_{3})\,,\hspace{1cm}\rho_{2}<\rho_{3}<\rho_{1}\\
                s(\rho_{3}-\rho_{1})\,,\hspace{1cm}\rho_{3}<\rho_{1}<\rho_{2}\\
                s(\rho_{3}-\rho_{2})\,,\hspace{1cm}\rho_{3}<\rho_{2}<\rho_{1}
                \end{array}
              \right.\,.
\end{equation}
The result \eqref{fieldlimit final1} can be rewritten in the following compact way
\begin{equation}\label{g=seis I's suma}
      \lim_{\alpha'\to 0}g^{1-\textrm{loop}}_{II}(s, t, u)=\frac{\kappa_{D}^{2}c(\gamma)}{\pi^{4-D/2}}\left[ I_{\gamma}(s, t)+I_{\gamma}(t, s)+I_{\gamma}(s, u)+I_{\gamma}(u, s)+I_{\gamma}(t, u)+I_{\gamma}(u, t)\right]
\end{equation}
where we have defined
\begin{equation}
    c(\gamma)=\frac{1}{4}\left( \frac{\pi}{4}\right)^{\gamma+1/2}\left[ \sin(\pi\gamma)\Gamma(\gamma+5/2)\right]^{-1}
\end{equation}
and the integral
\begin{eqnarray}\label{Igamma}
    I_{\gamma}(s, t)&=&t^{\gamma+1}\int_{0}^{1}dx\,\frac{(1-x)^{\gamma+1}}{sx-t(1-x)}\,.
\end{eqnarray}

The amplitude expressed in the form \eqref{g=seis I's suma} manifestly shows the invariance under permutations of the Mandelstam variables. Integral \eqref{Igamma} is the same as the one that appears in the case of $\mathcal{N}=4$ super Yang-Mills. This integral is finite for $4<D<8$. To study the limit $D\to 4$, it is convenient to go back to the dimensional regularization parameter $\epsilon=-2+D/2$. The dominant contribution of \eqref{Igamma} in a neighborhood of $\epsilon\simeq 0$ turns out to be
\begin{equation}
\label{I_epsilon}
     I_{-2-\epsilon}(s, t)=\frac{(st)^{-1}}{\epsilon}+\mathcal{O}(\epsilon^{0})\,.
\end{equation}
and, consequently, the amplitude \eqref{g=seis I's suma} reads
\begin{eqnarray}
     \lim_{\alpha'\to 0}g^{1-\textrm{loop}}_{II}(s, t, u)&=&\frac{2}{\epsilon^{2}}\frac{(s+t+u)}{stu}-\frac{1}{\epsilon}\left( \frac{s\log s+t\log t+u\log u}{stu}\right)+\mathcal{O}(\epsilon^{0})\,.\notag
\end{eqnarray}
The leading term above vanishes on-shell since $s+t+u=0$ and the logarithmic ones come from the subleading contribution in \eqref{I_epsilon}. From this, it can be seen that this reproduces the supergravity limit \cite{Green:1982sw,Schwarz:1982}. 

Notice that, contrary to what happens in Yang-Mills theory, where IR divergences manifest at order $\mathcal{O}(\epsilon^{-2})$, gravitational infrared infinities are softer since they start at $\mathcal{O}(\epsilon^{-1})$ order.


\section{Celestial amplitudes}\label{Celestial amplitudes}

\subsection{The celestial sphere}

In this section, we write the string amplitude of 4 gravitons at 1-loop in the conformal basis. As usual, we first need to express the amplitudes in terms of the energies for each external state, and the complex coordinates of their positions on celestial sphere at (null) infinity. For massless particles, their momentum reads,
\begin{equation}
\label{celestial_k}
    k^{\mu}_{i}=\eta_{i}\omega_{i} q^{\mu}_{i}(z_{i}, \bar{z}_{i})\,,
\end{equation}
where $\omega_{i}$ is the energy of the $i^{\text{th}}$ particle, and $\eta_{i}=\pm 1$ denotes whether the particle is incoming or outgoing. Here, $i=1,2,3,4$; $\mu=0,1,2,3$. $q^{\mu}_{i}$ is a null vector, $q^{2}_{i}=0$, parameterized in coordinates on the celestial sphere, $\{z_{i}, \bar{z}_{i}\}$; namely
\begin{equation}\label{momentumparametrization}
    q_{i}^{\mu}(z_{i}, \bar{z}_{i})=\frac{1}{2}\left( 1+|z_{i}|^{2},\;z_{i}+\bar{z}_{i},\;i(\bar{z}_{i}-z_{i}),\;1-|z_{i}|^{2}\right)\,.
\end{equation}
In this parameterization, the Mandelstam variables \eqref{Mandelstamstu2} read
\begin{eqnarray}\label{CelestialMandelstam}
    s&=&\eta_{1}\eta_{2}\omega_{1}\omega_{2}|z_{12}|^{2}=\eta_{3}\eta_{4}\omega_{3}\omega_{4}|z_{34}|^{2}\, \notag\\
    t&=&\eta_{2}\eta_{3}\omega_{2}\omega_{3}|z_{23}|^{2}=\eta_{1}\eta_{4}\omega_{1}\omega_{4}|z_{14}|^{2}\, \notag\\
    u&=&\eta_{1}\eta_{3}\omega_{1}\omega_{3}|z_{13}|^{2}=\eta_{2}\eta_{4}\omega_{2}\omega_{4}|z_{24}|^{2}\,,
\end{eqnarray}
where $z_{i j}=z_{i}-z_{j}$. In writing the second equalities above, one uses the total momentum conservation. The dimensionless ratio \eqref{cross-ratio 1} becomes the conformally invariant cross-ratio
\begin{eqnarray}\label{r=-s/t}
    r=-\frac{s}{t}=\frac{z_{12}z_{34}}{z_{23}z_{41}}\,,  \hspace{1cm}  1-r=\frac{z_{42}z_{13}}{z_{23}z_{41}}\,.
\end{eqnarray}
From the spinor-helicity formalism, we have
\begin{equation}\label{spinorhelicity celestial}
    \langle i j\rangle=\sqrt{\omega_{i}\omega_{j}}z_{i j}\,,\hspace{2cm}[i j]=-\sqrt{\omega_{i}\omega_{j}}\bar{z}_{i j}\,,
\end{equation}
where the Mandelstam variables read
\begin{eqnarray}
   s_{ij}&=&-\eta_i \eta_j \langle i j\rangle[i j]\,.
\end{eqnarray}
All this enables us to translate the usual amplitudes in momentum space into the conformal basis on the celestial sphere. See \cite{Strominger:2017zoo} and references therein and thereof; in particular, see \cite{Pasterski:2016qvg, Pasterski:2017kqt, Pasterski:2017ylz}. See also the reviews \cite{Pasterski:2021raf, Donnay:2020guq}. References \cite{Arkani-Hamed:2020gyp, Himwich:2020rro, Pate:2019lpp} are especially relevant for the discussion here.

Celestial amplitudes are obtained by computing the Mellin transform of the amplitudes in momentum space with respect to the frequencies $\omega_i$. This yields
\begin{eqnarray}\label{Mellin Transf}
   \mathcal{M}\left[ \mathcal{A}(\{\omega_{i},\vec{z}_{i}\})\right] &\equiv& \widetilde{\mathcal{A}}(\{\Delta_{i}, \vec{z}_{i}\})\notag\\
   &=&\int^{\infty}_{0}\prod^{n}_{i=1}\left(d\omega_{i}\, \omega_{i}^{\Delta_{i}-1}\right)\delta^{(4)}\left(\sum_{i=1}^{n}\eta_{i}\omega_{i}q^{\mu}_{i}\right)\,A(\{\omega_{i},\vec{z}_{i}\})\,,
\end{eqnarray}
where $\vec{z}_{i}=(z_{i},\bar{z}_{i})$. Also, hereafter we will denote the Mellin transform  of a momentum space amplitude as $\tilde{\mathcal{A}}=\mathcal{M}[\mathcal{A}]$. For the 2-to-2 scattering processes we are interested here, we use $\eta_1=\eta_2=-\eta_3=-\eta_4=1$. With these definitions, the total-momentum conserving delta function reads
\begin{eqnarray}\label{Dirac delta}
    \hspace{-0.5cm}\delta^{(4)}(\omega_{1}q_{1}+\omega_{2}q_{2}-\omega_{3}q_{3}-\omega_{4}q_{4})&=&\frac{4}{\omega_{4}|z_{14}|^{2}|z_{23}|^{2}}\,\delta(r-\bar{r})\,\delta\left(\omega_{1}-\frac{z_{24}\bar{z}_{34}}{z_{12}\bar{z}_{13}}\omega_{4} \right)\,\notag\\
    &&\times\,\delta\left(\omega_{2}-\frac{z_{14}\bar{z}_{34}}{z_{12}\bar{z}_{32}}\omega_{4} \right)\,\delta\left(\omega_{3}+\frac{z_{24}\bar{z}_{14}}{z_{23}\bar{z}_{13}}\omega_{4} \right)\,.
\end{eqnarray}
For external states reaching the 2-dimensional celestial sphere at null infinity, the conformal dimensions read $\Delta_{j}=1+i\lambda_{j}$. We also use the definition $\beta \equiv -\frac{i}{2}\sum_{j=1}^{4}\lambda_{j}\,,$ such that\footnote{See \cite{Arkani-Hamed:2020gyp} for the relevance of the analytic structure of celestial amplitudes in the $\beta$-complex plane.}
\begin{equation}\label{Deltasum}
    \sum_{j=1}^{4}\Delta_{j}=4-2\beta\,.
\end{equation}
According to the holographic proposal, the Mellin transform \eqref{Mellin Transf} maps scattering amplitudes of $n$ massless particles in momentum space into $n$-point correlation functions in a 2-dimensional CFT on the celestial sphere. The conformal weights of operators in the latter are $\left(h_i, \bar{h}_i\right)=\frac{1}{2}\left(\Delta_i+J_i, \Delta_i-J_i\right)$, with the 2D spins $\{J_{i}\}$ corresponding to the helicities of each of the external states. For the case $n=4$, the Mellin transform of the amplitude has the following form, which is fixed by the conformal symmetry,
\begin{eqnarray}
    \widetilde{\mathcal{A}}\left(\left\{\Delta_i, \vec{z}_i\right\}\right)&=&f(r, \bar{r}) \, \prod_{i<j}^4 z_{i j}^{\frac{h}{3}-h_i-h_j} \bar{z}_{i j}^{\frac{\bar{h}}{3}-\bar{h}_i-\bar{h}_j}\,.
\end{eqnarray}
The structure of the scattering process will be encoded in the function $f(r,\bar{r})$ that depends on the conformally invariant cross-ratio $r$ and the total scaling dimensions through $\beta$ \cite{Gonzalez:2020tpi,Arkani-Hamed:2020gyp}. For the particular case of an MHV amplitude of 4 gravitons, we have \cite{Stieberger:2018edy}
\begin{eqnarray}\label{pesos conformes}
h_1&=&-\frac{1}{2}+\frac{i}{2} \lambda_1  \,,\hspace{1cm} \bar{h}_1=\frac{3}{2}+\frac{i}{2} \lambda_1 \, \notag\\
h_2&=&-\frac{1}{2}+\frac{i}{2} \lambda_2  \,,\hspace{1cm} \bar{h}_2=\frac{3}{2}+\frac{i}{2} \lambda_2 \,,\\
h_3&=&\frac{3}{2}+\frac{i}{2} \lambda_3  \,,\hspace{1cm}\bar{h}_3=-\frac{1}{2}+\frac{i}{2} \lambda_3\,, \notag\\
h_4&=&\frac{3}{2}+\frac{i}{2} \lambda_4 \,,\hspace{1cm}\bar{h}_4=-\frac{1}{2}+\frac{i}{2} \lambda_4 \notag\,\,,
\end{eqnarray}
which yields $J_{1}=J_{2}=-2$, $J_{3}=J_{4}=2$ as expected.

In order to present our derivations more clearly, we will use the symbol $\stackrel{\delta}{=}$ to refer to the equations that hold after using the constraints on the energies given by the delta function \eqref{Dirac delta} along with the definition of the cross-ratio \eqref{r=-s/t}. For example, this allows us to rewrite the Mandelstam variables \eqref{CelestialMandelstam} as 
\begin{eqnarray}\label{celestialMandelstamdelta}
    s\stackrel{\delta}{=}\frac{z_{24}\bar{z}_{34}}{z_{12}\bar{z}_{13}}|z_{41}|^{2}r\omega_{4}^{2}\,,\hspace{1cm} t\stackrel{\delta}{=}-\frac{z_{24}\bar{z}_{34}}{z_{12}\bar{z}_{13}}|z_{41}|^{2}\omega_{4}^{2}\,,\hspace{1cm}  u\stackrel{\delta}{=}\frac{z_{24}\bar{z}_{34}}{z_{12}\bar{z}_{13}}|z_{41}|^{2}(1-r)\,\omega_{4}^{2}\,.
\end{eqnarray}


\subsection{Gravitons at tree-level}

In terms of the parametrization \eqref{celestial_k}, the tree-level 4-graviton MHV amplitude in pure gravity \eqref{tree-level 4-gravitonGR} takes the form 
\begin{eqnarray}\label{MHV gravitons celestial basis}
    A^{\textrm{tree }}_{GR}\left(r, \{\omega_{i}, \vec{z}_{i}\}\right)&=&\frac{\kappa^{2}_{10}}{4}\left(\frac{\omega_{1}\omega_{2}}{\omega_{3}\omega_{4}} \right)^{2}\left(\frac{z_{12}^{2}}{z^{2}_{34}} \right)^{2}\frac{\omega_{3}^{2}\omega_{4}}{\omega_{2}}\frac{|z_{34}|^{4}}{|z_{24}|^{2}}r\,\notag\\
    &\stackrel{\delta}{=} &\frac{\kappa^{2}_{10}}{4}\frac{z_{12}\bar{z}_{34}^{4}z_{24}z_{14}}{z_{34}^{2}\bar{z}_{13}\bar{z}_{12}\bar{z}_{23}}\frac{r^{2}\omega_{4}^{2}}{1-r}\,,
\end{eqnarray}
where in the first line we have also used \eqref{cross-ratio 1} to write $t=-s/r$ and \eqref{celestialMandelstamdelta} to simplify the dependence on ${z}_{i}$. Notice that, although we have explicity written the amplitude in $D=4$, we are including the 10-dimensional gravitational coupling $\kappa_{10}$; this will help us organize the celestial string loop expansion in a clearer form. Notice also that, when calculating the Mellin transforms, we have to include the factor $4/(\omega_{4}|z_{14}|^{2}|z_{23}|^{2})$ coming from the delta function \eqref{Dirac delta}.

The tree-level celestial amplitude of 4 gravitons for the heterotic string has been calculated in \cite{Stieberger:2018edy}. For the type II theory, it takes the form
\begin{eqnarray}\label{celestial tree strings}
\widetilde{\mathcal{A}}_{II}^{\textrm{tree}}\left( r, \beta,  \{\vec{z}_{i}\}\right)& =&2\kappa^{2}_{10}\left(\alpha'\right)^{\beta-1} \delta(r-\bar{r}) \Theta(r-1)\left(\prod_{i<j}^4 z_{i j}^{\frac{h}{3}-h_i-h_j} \bar{z}_{i j}^{\frac{\bar{h}}{3}-\bar{h}_i-\bar{h}_j}\right)\notag \\
&& \times r^{\frac{11-\beta}{3}}(r-1)^{-\frac{1+\beta}{3}}\mathcal{ G}(r, \beta)\,,
\end{eqnarray}
where
\begin{equation}
\mathcal{G}(r, \beta)= \int_0^{\infty} d\omega\,\omega^{3-\beta }r(r-1)\,g^{\textrm{tree}}_{II}(r\omega, -\omega, (1-r)\omega)\,.
\end{equation}

\subsection{Gravitons at one-loop}

Let us compute the Mellin transform of the 1-loop part of the MHV amplitude in type II superstring theory \eqref{massless amplitude type II}, namely
\begin{eqnarray}
  \tilde{ \mathcal{A}}^{1-\textrm{loop}}_{II}&=&\int^{\infty}_{0}\prod^{4}_{i=1}\left(d\omega_{i}\, \omega_{i}^{\Delta_{i}-1}\right)\delta^{(4)}\left(\sum_{j}^{4}\eta_{j}\omega_{j}q_{j}\right)stu \,A^{\textrm{tree }}_{GR}g^{1-\textrm{loop}}_{II}\, \notag\\   
    &\stackrel{\delta}{=} &-\kappa^{2}_{10}\, \frac{\bar{z}_{34}^{5}z^{3}_{24}|z_{14}|^{2}|z_{24}|^{2}|z_{34}|^{4}}{\bar{z}_{12}z_{34}^{2}\bar{z}_{13}^{3}|z_{23}|^{6}}\left(\frac{z_{24}\bar{z}_{34}}{z_{12}\bar{z}_{13}}\right)^{\Delta_{1}-1}\left(\frac{z_{14}\bar{z}_{34}}{z_{12}\bar{z}_{32}}\right)^{\Delta_{2}-1}\left(-\frac{z_{24}\bar{z}_{14}}{z_{23}\bar{z}_{13}} \right)^{\Delta_{3}-1}\notag\\
    &&\times \, \delta(r-\bar{r})\, \frac{r}{1-r}\int^{\infty}_{0}d\omega_{4}\,\omega_{4}^{\Delta_{1}+\Delta_{2}+\Delta_{3}+\Delta_{4}+3}g^{1-\textrm{loop}}_{II}(\{\omega_{4}\})\,,
\end{eqnarray}
where we have used \eqref{celestialMandelstamdelta} and \eqref{MHV gravitons celestial basis},  
\begin{eqnarray}
    stu=\frac{\omega_{2}\omega_{3}^{2}\omega_{4}^{3}}{r}|z_{34}|^{4}|z_{24}|^{2}\stackrel{\delta}{=}-\frac{z^{2}_{24}\bar{z}_{14}\bar{z}_{34}|z_{14}|^{2}|z_{24}|^{2}|z_{34}|^{4}}{z_{23}\bar{z}_{13}^{2}z_{12}|z_{23}|^{2}}\frac{\omega^{6}_{4}}{r}\,.
\end{eqnarray}
Function $g^{1-\textrm{loop}}_{II}(\{\omega_{i}\})$ is given by
\begin{eqnarray}
&&    g^{1-\textrm{loop}}_{II}(\{\omega_{4}\})=\frac{\kappa^{2}_{10}}{\alpha'}\int_{\mathcal{F}} \frac{d\tau d\bar{\tau}}{\left(\textrm{Im}(\tau)\right)^{5}}[F_{2}(a, \tau)]^{10-D}\, \times  \nonumber\\
      && \ \ \ \int\prod^{3}_{l=1}d^{2}\nu_{l}\left(\frac{\chi(\nu_{1 2},\tau)\chi(\nu_{3 4}, \tau)}{\chi(\nu_{1 3}, \tau)\chi(\nu_{2 4},\tau)}\right)^{-\frac{\alpha'}{2}\frac{z_{24}\bar{z}_{34}}{z_{12}\bar{z}_{13}}|z_{41}|^{2}r\omega_{4}^{2}} \left(\frac{\chi(\nu_{2 3},\tau)\chi(\nu_{1 4}, \tau)}{\chi(\nu_{1 3},\tau)\chi(\nu_{2 4},\tau)}\right)^{\frac{\alpha'}{2}\frac{z_{24}\bar{z}_{34}}{z_{12}\bar{z}_{13}}|z_{41}|^{2}\omega_{4}^{2}}\nonumber\,
\end{eqnarray}
where we have used \eqref{celestialMandelstamdelta}.
 Defining the new variable
\begin{equation}
    \omega=\frac{1}{2}\frac{z_{24}\bar{z}_{34}}{z_{12}\bar{z}_{13}}|z_{41}|^{2}\omega_{4}^{2}\,,
\end{equation}
which satisfies $ \omega=-\frac{1}{2}t\geq 0$
in the physical region, the celestial amplitude can be written as follows
\begin{eqnarray}
    \tilde{ \mathcal{A}}^{1-\textrm{loop}}_{II} &=&\kappa^{2}_{10}2^{3-\beta}\frac{\bar{z}_{34}^{5}z^{3}_{24}|z_{14}|^{2}|z_{24}|^{2}|z_{34}|^{4}}{\bar{z}_{12}z_{34}^{2}\bar{z}_{13}^{3}|z_{23}|^{6}}\left(\frac{z_{24}\bar{z}_{34}}{z_{12}\bar{z}_{13}}\right)^{\Delta_{1}-1}\left(\frac{z_{14}\bar{z}_{34}}{z_{12}\bar{z}_{32}}\right)^{\Delta_{2}-1}\left(-\frac{z_{24}\bar{z}_{14}}{z_{23}\bar{z}_{13}} \right)^{\Delta_{3}-1}\nonumber\\
    &&\times\left(\frac{z_{12}\bar{z}_{13}}{z_{24}\bar{z}_{34}|z_{41}|^{2}}\right)^{4-\beta}  \delta(r-\bar{r})\, \frac{r}{1-r}\, \mathcal{J}(r,\beta)\,,
\end{eqnarray}
where we have used the definition of $\beta$, \eqref{Deltasum} and we have denoted
\begin{eqnarray}\label{J function 0}
    &&\mathcal{J}(r,\beta)=\frac{\kappa^{2}_{10}}{\alpha'} \int_{\mathcal{F}} \frac{d\tau d\bar{\tau}}{\left(\textrm{Im}(\tau)\right)^{5}}\, [F_{2}(a, \tau)]^{10-D}\,\\
    &&\ \ \ \ \ \times\, \int^{\infty}_{0}d\omega\,\omega^{-\beta+3}\int\prod^{3}_{l=1}d^{2}\nu_{l}\left(\frac{\chi(\nu_{1 2},\tau)\chi(\nu_{3 4}, \tau)}{\chi(\nu_{1 3}, \tau)\chi(\nu_{2 4},\tau)}\right)^{-\alpha'r\omega}\left(\frac{\chi(\nu_{2 3},\tau)\chi(\nu_{1 4}, \tau)}{\chi(\nu_{1 3},\tau)\chi(\nu_{2 4},\tau)}\right)^{\alpha'\omega}\,.\notag
\end{eqnarray}
We can further reduce the expression by defining
\begin{equation}
    X\equiv \log\left(\frac{\chi(\nu_{1 2},\tau)\chi(\nu_{3 4}, \tau)}{\chi(\nu_{1 3}, \tau)\chi(\nu_{2 4},\tau)} \right)\,,\hspace{1cm}Y\equiv \log\left(\frac{\chi(\nu_{2 3},\tau)\chi(\nu_{1 4}, \tau)}{\chi(\nu_{1 3},\tau)\chi(\nu_{2 4},\tau)}\right)\,,
\end{equation}
and conveniently managing the ${z}_{i}$-dependent factors. With this, we arrive at
\begin{multline}
\label{casi celestial amplitude}
\tilde{\mathcal{A}}^{1-\textrm{loop}}_{II}\left(  r, \beta, \{\vec{z}_{i}\}\right) = \kappa^{2}_{10}\,2^{3-\beta}\,\delta(r-\bar{r})\Theta(r-1)\\\times r^{\frac{14-\beta}{3}}(1-r)^{\frac{2-\beta}{3}}
\mathcal{J}(r, \beta)\,\prod_{i<j}^4 z_{i j}^{\frac{h}{3}-h_i-h_j} \bar{z}_{i j}^{\frac{\bar{h}}{3}-\bar{h}_i-\bar{h}_j}\,,
\end{multline}
where we have simplified the expression using the results in the Appendix. $\Theta(x)$ is the Heaviside step function that imposes the physical condition $r>1$ at the very end of the computation. We will omit the function $\Theta(r-1)$ in what follows. We have also exchanged the order of the integrals in $\nu_{l}$ and $\omega$, now having
\begin{equation}
   \mathcal{J}(r, \beta)= \frac{\kappa^{2}_{10}}{\alpha'} \int_{\mathcal{F}} \frac{d\tau d\bar{\tau}}{\left(\textrm{Im}(\tau)\right)^{5}}\, [F_{2}(a, \tau)]^{10-D}\,\int\prod^{3}_{l=1}d^{2}\nu_{l}\int^{\infty}_{0}d\omega\,\omega^{-\beta+3}e^{-\alpha'\omega( rX-Y)}\,.
\end{equation}
from where we see that it is possible to fully compute the Mellin transform. Introducing the change of variables
\begin{equation}
    \eta=\alpha'\omega(rX-Y)\,,
\end{equation}
we have
\begin{eqnarray}
   \int^{\infty}_{0}d\omega\,\omega^{-\beta+3}e^{-\alpha'\omega( rX-Y)}=(\alpha')^{\beta-4}(rX-Y)^{\beta-4}\Gamma(4-\beta)\,.
\end{eqnarray}
This enables us to write the full dependence in $\alpha'$ encapsulated as an overall factor, namely
\begin{equation}\label{J function}
   \mathcal{J}(r, \beta) =\kappa_{10}^{2} (\alpha')^{\beta-5} \Gamma(4-\beta)\int_{\mathcal{F}} \frac{d\tau d\bar{\tau}}{\left(\textrm{Im}(\tau)\right)^{5}}\,[F_{2}(a, \tau)]^{10-D}\int\prod^{3}_{l=1}d^{2}\nu_{l}\, (rX-Y)^{\beta-4}\,.
\end{equation}
This peculiar dependence on $\alpha'$ that the string amplitudes exhibit in the conformal basis had been observed in other processes at tree and 1-loop levels \cite{Stieberger:2018edy, Castiblanco:2024hnq, Donnay:2023kvm, Bockisch:2024bia} and it is a distinctive feature of celestial holography that is not displayed in \emph{Carrollian} string theory amplitudes \cite{Stieberger:2024shv}. 

Putting all these results together and adding the tree-level contribution \eqref{celestial tree strings}, we obtain that the expression for the expansion in the celestial basis, through 1-loop, reads
\begin{equation}
    \tilde{\mathcal{A}}_{II}=\kappa_{10}^{2}(\alpha')^{\beta-1}\left( \tilde{\mathcal{A}}^{(0)}_{II}+\frac{\kappa_{10}^{2}}{(\alpha')^{4}}\tilde{\mathcal{A}}^{(1)}_{II}+\cdots\right)\,,
\end{equation}
with $\kappa_{10}^{2}/(\alpha')^{4}$ being the dimensionless parameter that organizes the loop expansion in the string celestial amplitudes. Here, we have also defined the dimensionless quantities $\tilde{\mathcal{A}}^{(0)}_{II}=\kappa^{-2}_{10}(\alpha')^{1-\beta}\tilde{\mathcal{A}}^{\rm tree}_{II}$ and $\tilde{\mathcal{A}}^{(1)}_{II}=\kappa^{-2}_{10}(\alpha')^{5-\beta}\tilde{\mathcal{A}}^{1-\rm loop}_{II}$.


\subsection{Field theory limit of celestial strings}

As mentioned earlier, the field theory limit is controlled by the region $\textrm{Im}\tau\to\infty$ of the moduli space, where we have \cite{ Green:1982sw, Schwarz:1982}
\begin{equation}
     F_{2}(a, \tau)\simeq a\left( \textrm{Im}\tau\right)^{1/2}\,.
\end{equation}
In order to obtain the dominant contribution of \eqref{J function}, we need to keep in mind that the function $\mathcal{J}(r, \beta)$ involves an integral over the fundamental domain of the moduli space \eqref{FundamentalDom. torus}, such that if we use $  \sin (\pi\nu_{ji})\simeq\frac{i}{2}{e^{- i\pi\nu_{ji}}}$ for $\textrm{Im}\nu_{ij}\to \infty$, we obtain from \eqref{chifunction} that in the limit ${\textrm{Im}\tau\to \infty}$,
\begin{eqnarray}      \chi_{i j}&\simeq&\exp\left\{ -\frac{\pi(\textrm{Im}\nu_{ij})^{2}}{\textrm{Im}\tau}+\pi\textrm{Im}(\nu_{ ji})\right\}\,,
\end{eqnarray}
such that
\begin{eqnarray}\label{limit rX-Y v1}
(rX-Y)&\simeq &  -\frac{\pi}{\textrm{Im}\tau} r\left[(\textrm{Im}\nu_{12})^{2}+(\textrm{Im}\nu_{34})^{2}-(\textrm{Im}\nu_{13})^{2}-(\textrm{Im}\nu_{24})^{2}\right]+2\pi r\,\textrm{Im}\nu_{2 3}\notag\\
&& +\frac{\pi}{\textrm{Im}\tau} \left[(\textrm{Im}\nu_{23})^{2}+(\textrm{Im}\nu_{14})^{2}-(\textrm{Im}\nu_{13})^{2}-(\textrm{Im}\nu_{24})^{2}\right]\,.
\end{eqnarray}
Notice that the integrand of \eqref{J function} only depends on the imaginary part of $\tau$ and the locations of the vertex operators $\nu_i$. Therefore, we can integrate directly over the real part, so that, by defining the variables
\begin{eqnarray}\label{rho coordinates}
   \rho_{i} &=&\frac{\textrm{Im}\nu_{i}}{\textrm{Im}\tau}\,,\hspace{1cm}\textrm{Im}\tau\equiv T\,,
\end{eqnarray}
we get
\begin{eqnarray}
    \int_{\mathcal{F}}\frac{d\tau d\bar{\tau}}{\left( \textrm{Im}\tau\right)^{5}}\int\left(\prod^{3}_{l=1}d^{2}\nu_{l}\right)=\int_{1}^{\infty}\frac{\textrm{Im}(d\tau) }{\left( \textrm{Im}\tau\right)^{5}}\int_{0}^{\infty}\prod^{3}_{l=1}\textrm{Im}(d\nu_{l})=\int_{1}^{\infty}\frac{dT}{T^{2}} \int_{0}^{1}[d\rho]\;+\;\textrm{perm}\,.\nonumber
\end{eqnarray}
The permutations come from the definitions of variables \eqref{rho coordinates} when integrating on the fundamental domain. For example, for the particular ordering $0\leq \rho_{1}< \rho_{2}< \rho_{3}\leq \rho_{4}\equiv 1$, in the limit $\textrm{Im}\tau\to\infty$ expression \eqref{limit rX-Y v1} is rewritten as follows
\begin{eqnarray}
\hspace{-1cm}rX-Y&\simeq &2\pi\,T\left[(1-r)\rho_{1}\rho_{3} +\rho_{2}+r\rho_{1}\rho_{2}-\rho_{2}\rho_{3}-\rho_{1}\right]\,. 
\end{eqnarray}
Eventually, in the $\textrm{Im}\tau\to\infty$ limit, the integral \eqref{J function} becomes
\begin{eqnarray}
 \mathcal{J}(r, \beta) &=&\kappa_{10}^{2}(\alpha')^{\beta-5} a^{10-D}(2\pi\,)^{\beta-4} \Gamma(4-\beta)\int_{1}^{\infty}\frac{dT}{T} \,T^{\,\beta-D/2}\\
   &&\times\left(\int_{0}^{1}[d\rho]\,\left[(1-r)\rho_{1}\rho_{3} +\rho_{2}+r\rho_{1}\rho_{2}-\rho_{2}\rho_{3}-\rho_{1}\right]^{\beta-4}+\;\textrm{perm}\right)\notag\,.
\end{eqnarray}
The power of $a$ appearing in this formula, which comes from $F_{2}(a, \tau)$, is absorbed in the identification of the couplings. More precisely, we have $\kappa_{D}^{2}=\kappa^{2}_{10}R^{D-10}$ and $a=\sqrt{\alpha'}/R$, so that
\begin{equation}
 \kappa^{2}_{10}(\alpha')^{\beta-5}a^{10-D}=\kappa_{D}^{2}(\alpha')^{\beta-D/2}\,.   
\end{equation}
By defining $\eta=(\alpha'T)^{-1}$, we can absorb the dependence on $\alpha'$ into an integration limit
\begin{eqnarray}\label{TTTTT}
    (\alpha')^{\beta-D/2} \int_{1}^{\infty}\frac{dT}{T} \,T^{\,\beta-D/2}&=&\int^{1/\alpha'}_{0} \frac{d\eta}{\eta}\,\eta^{\frac{i}{2}\lambda+D/2}\,.
\end{eqnarray}
where we use $\beta=-\frac{i}{2}\lambda$, with $\lambda=\sum_{i=1}^4\lambda_i$; here $D=4$. With all this, we find that the $\alpha'\to 0$ limit is
\begin{eqnarray}\label{alpha'=0 J function}
  \lim_{\alpha'\to 0} \mathcal{J}(r, \beta) &=&\kappa_{D}^{2} (2\pi\,)^{\beta-4}\Gamma(4-\beta)\,\mathcal{I}(\lambda-iD)\notag\\
   &\times&\left(\int_{0}^{1}[d\rho]\,\left[(1-r)\rho_{1}\rho_{3} +\rho_{2}+r\rho_{1}\rho_{2}-\rho_{2}\rho_{3}-\rho_{1}\right]^{\beta-4}+\;\textrm{perm}\right)\notag\,,
\end{eqnarray}
with
\begin{equation}\label{I function}
    \mathcal{I}(x)=\int_{0}^{\infty} {d\eta}\, \eta^{\frac{i}{2}x-1}\,.
\end{equation}
Notice that all the $\alpha'$ dependence dissappeared, as it should, since we are arriving at a field theory amplitude.

The integral (\ref{I function}) for $x \in \mathbb{R}$ realizes a delta function $\mathcal{I}(x)=4\pi \delta(x)$. The extension of this kernel for non-real $x$ has previously appeared in the context of celestial holography, cf. \cite{Puhm:2019zbl, Gonzalez:2020tpi}, and has been carefully analyzed in \cite{Borji:2024pvg} where it was shown that, for non-real $x$, the divergent integral $\mathcal{I}(\lambda-iD)$ is again distributional, consistent with the fact that scattering amplitudes are indeed tempered distributions \cite{Borji:2024pvg}.

Putting all together, the $\alpha'\to 0$ limit of the 1-loop closed string celestial amplitude reads
\begin{eqnarray}\label{field limit celestial amplitude}
    \lim_{\alpha'\to 0}\mathcal{M}\left[ \mathcal{A}^{1-\textrm{loop}}_{II} \right] &=&\kappa^{4}_{D}(2\pi)^{\beta-4}\,2^{3-\beta}\Gamma(4-\beta)\,\mathcal{I}(\lambda-iD)\notag\\
    &&\times\, \delta(r-\bar{r})\,r^{\frac{14-\beta}{3}}(1-r)^{\frac{2-\beta}{3}}\, \prod_{i<j}^4 z_{i j}^{\frac{h}{3}-h_i-h_j} \bar{z}_{i j}^{\frac{\bar{h}}{3}-\bar{h}_i-\bar{h}_j}\notag\\
   &&\times \left(\int_{0}^{1}[d\rho]\,\left[(1-r)\rho_{1}\rho_{3} +r\rho_{1}\rho_{2}-\rho_{2}\rho_{3}-(\rho_{1}-\rho_{2})\right]^{\beta-4}+\;\textrm{perm}\right)\notag\,.\\
\end{eqnarray}
Recall that we are denoting the celestial correlator of an amplitude $\mathcal{A}$, as $\tilde{\mathcal{A}}=\mathcal{M}[\mathcal{A}]$.

In the next subsection we will show that this expression is exactly recovered by a different procedure, namely, by directly performing the Mellin transforms of the momentum space amplitude obtained in quantum field theory. In other words, in spite of the energy mixing, the Mellin transform for the string amplitude is found to commute with the $\alpha'\to 0$ limit, when properly defined.


\subsection{Celestial supergravity}

Let us now compute the celestial correlator corresponding to the 1-loop, four graviton amplitude in quantum field theory and investigate how it reproduces our result \eqref{field limit celestial amplitude}. The full celestial amplitude has the form
\begin{eqnarray}
&&   \mathcal{M}\left[\lim_{\alpha'\to 0}\mathcal{A}_{II}^{1-\textrm{loop}}\right]=\int^{\infty}_{0}\prod^{4}_{i=1}\left(d\omega_{i}\, \omega_{i}^{\Delta_{i}-1}\right)\delta^{(4)}\left(\sum_{j}^{4}\eta_{j}\omega_{j}q_{j}\right)stu \,A^{\textrm{tree }}_{GR}\left(  \lim_{\alpha'\to 0}g^{1-\textrm{loop}}_{II}\right)\, \notag\\      & & \ \ \ \ \ \ \ \ \ \ \ =\, \kappa^{2}_{D}\frac{\bar{z}_{34}^{5}z^{3}_{24}|z_{14}|^{2}|z_{24}|^{2}|z_{34}|^{4}}{\bar{z}_{12}z_{34}^{2}\bar{z}_{13}^{3}|z_{23}|^{6}}\left(\frac{z_{24}\bar{z}_{34}}{z_{12}\bar{z}_{13}}\right)^{\Delta_{1}-1}\left(\frac{z_{14}\bar{z}_{34}}{z_{12}\bar{z}_{32}}\right)^{\Delta_{2}-1}\left(-\frac{z_{24}\bar{z}_{14}}{z_{23}\bar{z}_{13}} \right)^{\Delta_{3}-1}\notag\\
    &&\  \ \  \ \ \ \ \  \ \ \ \ \ \times\, \delta(r-\bar{r})\,\frac{r}{1-r}\,\int^{\infty}_{0}d\omega_{4}\,\omega_{4}^{-2\beta+7}\left(  \lim_{\alpha'\to 0}g^{1-\textrm{loop}}_{II}(\omega_{4}, \{\vec{z}_{i}\})\right)\,,
\end{eqnarray}
which, in particular, contains the form \eqref{fieldlimit final1} expressed in the celestial angular coordinates, 
\begin{eqnarray}
 & & \lim_{\alpha'\to 0}g^{1-\textrm{loop}}_{II}(\omega_{4}, \{\vec{z}_{i}\})=\kappa_{D}^{2}\pi^{D/2-4}\Gamma(4-D/2)\left(\frac{z_{24}\bar{z}_{34}}{z_{12}\bar{z}_{13}}|z_{41}|^{2}\right)^{-4+D/2}(\omega_{4}^{2})^{-4+D/2}\nonumber \\
     && \ \ \ \ \ \ \ \ \ \ \times\, \left(\int _{0}^{1}[d\rho][(1-r)\rho_{3}\rho_{1}+r\rho_{1}\rho_{2}-\rho_{2}\rho_{3}-(\rho_{1}-\rho_{2})]^{-4+D/2}+\textrm{perm}\right)
\end{eqnarray}
Considering the variable $\omega_{4}=\omega^{1/2}\geq 0$, the integral over the energy $\omega_{4}$ becomes
\begin{equation}
    \int^{\infty}_{0}d\omega_{4}\,\omega_{4}^{D-2\beta -1 }=\frac{1}{2}\int^{\infty}_{0}\frac{d\omega}{\omega}\omega^{\frac{i}{2}\lambda+\frac{D}{2}}\equiv \frac{1}{2}\mathcal{I}(\lambda-iD)\,,
\end{equation}
and the Mellin transform of the field amplitude becomes
\begin{eqnarray}
   & & \mathcal{M}\left[\lim_{\alpha'\to 0}\mathcal{A}_{II}^{1-\textrm{loop}}\right]= \kappa^{4}_{D}(2\pi)^{D/2-4}\,2^{3-D/2}\Gamma(4-D/2)\,\mathcal{I}(\lambda-iD)\,\delta(r-\bar{r})\, \frac{r}{1-r}\notag\\
   &&\ \ \ \ \ \times\, \frac{\bar{z}_{34}^{5}z^{3}_{24}|z_{14}|^{2}|z_{24}|^{2}|z_{34}|^{4}}{\bar{z}_{12}z_{34}^{2}\bar{z}_{13}^{3}|z_{23}|^{6}}\left(\frac{z_{24}\bar{z}_{34}}{z_{12}\bar{z}_{13}}\right)^{i\lambda_{1}}\left(\frac{z_{14}\bar{z}_{34}}{z_{12}\bar{z}_{32}}\right)^{i\lambda_{2}}\left(-\frac{z_{24}\bar{z}_{14}}{z_{23}\bar{z}_{13}} \right)^{i\lambda_{3}}\left(\frac{z_{12}\bar{z}_{13}}{z_{24}\bar{z}_{34}|z_{41}|^{2}}\right)^{4-D/2}\notag\\
   &&\ \ \ \ \ \times\, \left(\int _{0}^{1}[d\rho][(1-r)\rho_{3}\rho_{1}+r\rho_{1}\rho_{2}-\rho_{2}\rho_{3}-(\rho_{1}-\rho_{2})]^{-4+D/2}+\textrm{perm}\right)\,.
\end{eqnarray}
Finally, in order to facilitate the comparison with \eqref{field limit celestial amplitude}, we can rewrite the ${z}_{i}$-dependent factors using the result \eqref{z factors} in the Appendix. Notice also the relations $\frac D2 =2= -\frac{i}{2}\sum_{i=1}^4\lambda_{i}=\beta$ imposed by the $\mathcal{I}(\lambda-iD)$ distribution. This yields the final result
\begin{eqnarray}\label{mellin of field limit}
    & & \mathcal{M}\left[\lim_{\alpha'\to 0}\mathcal{A}_{II}^{1-\textrm{loop}}\right]=\kappa^{4}_{D}  (2\pi)^{D/2-4}\,2^{3-D/2}\,\Gamma(4-D/2)\,\mathcal{I}(\lambda-iD) \notag\\
    && \ \ \ \ \ \ \ \ \ \times \delta(r-\bar{r})\,r^{\frac{14-D/2}{3}}(1-r)^{\frac{2-D/2}{3}}\, \prod_{i<j}^4 z_{i j}^{\frac{h}{3}-h_i-h_j} \bar{z}_{i j}^{\frac{\bar{h}}{3}-\bar{h}_i-\bar{h}_j} \\
   && \ \ \ \ \times \left(\int _{0}^{1}[d\rho][(1-r)\rho_{3}\rho_{1}+r\rho_{1}\rho_{2}-\rho_{2}\rho_{3}-(\rho_{1}-\rho_{2})]^{D/2-4}+\textrm{perm}\right)\,.\notag
\end{eqnarray}
which, after evaluating the distributional function (\ref{I function}), is found to exactly reproduce \eqref{field limit celestial amplitude}. Notice that, while the exponent in the third line of (\ref{mellin of field limit}) is $\frac D2 -4 $, the exponent in the third line of (\ref{field limit celestial amplitude}) is $\beta -4$, and so it is the presence of the overall factor $\mathcal{I}(\lambda-iD)$ which, for being of distributional nature \cite{Borji:2024pvg} and enforcing $\beta=\frac D2$ what makes the two expressions to match.

Note also that, instead of performing the Mellin transforms directly on the field theory expression \eqref{1-loop 4-gravitonN=8} (that one usually obtains from, for instance, Feynman diagrams), we do it in the \emph{worldline} formalism which is more direct when arriving at particle amplitudes from string theoretic computations. However, if we perform the integral above, double poles in the $\beta$-complex plane will indeed appear, in agreement with \cite{Arkani-Hamed:2020gyp} whose authors pointed out that these double poles are a consequence of the logarithms in the Mandelstam invariants that appear at 1-loop in $D=4$.

In conclusion, we observe that, at 1-loop, the Mellin transform commutes with the field theory limit for arbitrary values of the Mandelstam ratio $r$. At tree-level, the following explanation was suggested for the fact that the quantum field theory limit is recovered in the kinematic limit of large
$r$: it turns out that large $r$ corresponds to the forward scattering at $\theta =0$, where the process is
dominated by the exchanges of massless particles \cite{Stieberger:2018edy}. In the case of the 1-loop amplitude, we have shown that one recovers the field theory result in the limit $\alpha ' \to 0$ without requiring to take the large $r$ limit. This is because in the 1-loop calculation one deals with the integral in the moduli space of the genus-1 surface, which allows to redefine the integration range of the modular parameter and thus absorb the dependence on $\alpha '$ to match the entire domain of integration of the Schwinger parameters in the field theory. This is what has been done in \cite{Donnay:2023kvm} for the case of the gluons on the annulus, where the 1-loop gluon amplitudes expressed in the worldline formalism were recovered as $\alpha ' $ approaches zero. The computation for gravitons we have presented here works in a similar way; this can be seen from (\ref{TTTTT}), where the range of parameter in $\eta = (\alpha ' \textrm{Im}\tau)^{-1}$ allows us to absorb the dependence on $\alpha '$ in a way that it disappears from the leading contribution in the $\alpha ' \to 0 $ limit.

One interesting question that one might ask is how the differences found between tree-level and 1-loop computation in the celestial basis are compatible with the open-closed string duality, according to which certain 1-loop processes in open string theory can be analogously described in terms of classical string diagrams in closed string theory. Since in the limit $\alpha ' \to 0$ the celestial amplitudes at tree-level and 1-loop behave differently with respect to the parameter $r$, it is interesting to ask how the open-closed string duality works in that basis. Answering this question would require a better understanding of the tree-level celestial amplitude and of the interplay is between the moduli space and the classical limit in that case. This requires further study. As far as the 1-loop calculation we carried out here is concerned, our result correctly reproduces that of field theory in the right limit.

\section{Discussion}

In this paper we have studied 1-loop closed string scattering amplitudes in the conformal basis, focusing on the case of 4-graviton processes in the type II superstring. 

We have shown that, as previously observed for gluon amplitudes at tree- and 1-loop levels, the dependence on $\alpha'$ in the celestial basis factorizes out. This is a peculiar feature of celestial string amplitudes that is not observed in other attempts to formulate a flat space holography in the context of string theory \cite{Stieberger:2024shv}. As opposed to the traditional momentum basis, this factorization precludes us from simply taking $\alpha' \to 0$ to arrive at the field theory limit of the corresponding string amplitude. Instead, one needs to perform a careful analysis of the regions in the string moduli space that dominate in this limit, which is particularly more involved at 1-loop as compared to the tree level case. When properly defined, the $\alpha'\to 0 $ limit of the closed string 1-loop amplitudes commutes with the Mellin transforms. This yields that the field theory limit of the celestial string amplitude \eqref{field limit celestial amplitude} matches with the celestial field theory result \eqref{mellin of field limit} for all values of the Mandelstam ratio $r=-s/t$. This had already been observed in the case of 1-loop amplitudes for gluons in open string theory, as opposed to the tree level case, and here we verify that the same phenomenon occurs also for closed string gravitons at 1-loop. In this sense, this work can be considered as a continuation of the line of research initiated in \cite{Stieberger:2018edy, Donnay:2023kvm, Bockisch:2024bia, Castiblanco:2024hnq}, dedicated to investigating celestial holography within the context of perturbative string theory. There are several questions that arise from this series of works. One is how the mapping between the worldsheet and the celestial CFT observed in \cite{Stieberger:2018edy} and further explored in \cite{Castiblanco:2024hnq, eldehoy} generalizes to the closed string case. It would also be interesting to
study the open-closed duality in relation to the celestial basis, as well as to investigate the double copy in this context, cf. \cite{eldehoy}. We hope to address these questions in the near future.

\hspace{1cm}

\section*{Acknowledgments}

G.G. thanks Crist\'obal Corral and the Universidad Adolfo Ib\'a\~nez for the hospitality during his stay, where part of this work was done. He also thanks Massimo Porrati for discussions. The work of A.F.C.G. is supported by FONDECYT Regular grant 1201145 and Anillo Grant ANID/ACT210100. Y.P.C. is supported by Beca Doctorado Nacional (ANID) 2024 Scholarship N$^{\circ }$ 21242359.  F.R. is supported by FONDECYT Regular grants 1221920 and 1230853, and Anillo Grant ANID/ACT210100. F.R. would also like thank Hern\'an Gonz\'alez, Andrea Puhm, Stephan Stieberger and Tomasz Taylor for valuable discussions and especially to Stephan Stieberger for comments on the draft. G.G. and F.R. thank the organizers of the Annual Meeting of the Simons Collaboration on Celestial Holography for the invitation and the hospitality at the Flatiron Institute, where this work was completed.


\appendix

\section{Conformal factor }\label{z's factors in correlation function }

Conformal symmetry forces the 4-point correlation function on the celestial sphere to have the conformally covariant form
\begin{equation}
   \widetilde{A}(\{\Delta, \vec{z}_{i}\})={f}(r, \tilde{r})\, \prod_{i<j}^4 z_{i j}^{\frac{h}{3}-h_i-h_j} \bar{z}_{i j}^{\frac{\bar{h}}{3}-\bar{h}_i-\bar{h}_j}\,,
\end{equation}
where $(h_{i}, \bar{h}_{i})$ correspond to the conformal weights of the dual fields in the celestial CFT, given by \eqref{pesos conformes}, with $h=\sum_{i}h_{i}$. Expanding all these factors, one gets
\begin{eqnarray}\label{expansion z}
    \prod_{i<j}^4 z_{i j}^{\frac{h}{3}-h_i-h_j} \bar{z}_{i j}^{\frac{\bar{h}}{3}-\bar{h}_i-\bar{h}_j}&=&
    =\left(\frac{z_{23}^{1/2}\bar{z}_{23}^{1/2}z_{24}^{1/2}\bar{z}_{24}^{1/2}z_{34}^{1/2}\bar{z}_{34}^{1/2}}{z_{12}\bar{z}_{12}z_{13}\bar{z}_{13}z_{14}\bar{z}_{14}} \right)^{\frac{i\lambda_{1}}{3}} \left( \frac{z_{13}^{1/2}\bar{z}_{13}^{1/2}z_{14}^{1/2}\bar{z}_{14}^{1/2}z_{34}^{1/2}\bar{z}_{34}^{1/2}}{z_{12}\bar{z}_{12}z_{23}\bar{z}_{23}z_{24}\bar{z}_{24}}\right)^{\frac{i\lambda_{2}}{3}}  \notag\\
   &&\times\left(\frac{z_{12}^{1/2}\bar{z}_{12}^{1/2}z_{14}^{1/2}\bar{z}_{14}^{1/2}z_{24}^{1/2}\bar{z}_{24}^{1/2}}{z_{13}\bar{z}_{13}z_{23}\bar{z}_{23}z_{34}\bar{z}_{34}} \right)^{\frac{i\lambda_{3}}{3}} \left( \frac{z_{12}^{1/2}\bar{z}_{12}^{1/2}z_{13}^{1/2}\bar{z}_{13}^{1/2}z_{23}^{1/2}\bar{z}_{23}^{1/2}}{z_{14}\bar{z}_{14}z_{24}\bar{z}_{24}z_{34}\bar{z}_{34}}\right)^{\frac{i\lambda_{4}}{3}}  \notag\\
    && \times \left( \frac{z_{12}^{5}\bar{z}_{34}^{5}}{z_{13}z_{14}z_{23}z_{24}z_{34}^{7}\bar{z}_{12}^{7}\bar{z}_{13}\bar{z}_{14}\bar{z}_{23}\bar{z}_{24}}\right)^{1/3}  \,.
\end{eqnarray}
By repeated use of the $r=\bar{r}$ condition imposed by the delta function \eqref{Dirac delta} (planarity of the general 4-point amplitude), and \eqref{r=-s/t}, these factors can be written as
\begin{eqnarray}
    \left(\frac{z_{23}^{1/2}\bar{z}_{23}^{1/2}z_{24}^{1/2}\bar{z}_{24}^{1/2}z_{34}^{1/2}\bar{z}_{34}^{1/2}}{z_{12}\bar{z}_{12}z_{13}\bar{z}_{13}z_{14}\bar{z}_{14}} \right)^{\frac{i\lambda_{1}}{3}}&=&r^{-\frac{i\lambda_{1}}{6}}(1-r)^{-\frac{i\lambda_{1}}{6}}\left(\frac{z_{24}\bar{z}_{34}}{z_{12}\bar{z}_{13}}\right)^{i\lambda_{1}}\left(\frac{z_{12}\bar{z}_{13}}{z_{24}\bar{z}_{34}|z_{14}|^{2}}\right)^{\frac{i\lambda_{1}}{2}}\, \notag\\
     \left( \frac{z_{13}^{1/2}\bar{z}_{13}^{1/2}z_{14}^{1/2}\bar{z}_{14}^{1/2}z_{34}^{1/2}\bar{z}_{34}^{1/2}}{z_{12}\bar{z}_{12}z_{23}\bar{z}_{23}z_{24}\bar{z}_{24}}\right)^{\frac{i\lambda_{2}}{3}} &=&r^{-\frac{i\lambda_{2}}{6}}(1-r)^{-\frac{i\lambda_{2}}{6}}\left(\frac{z_{14}\bar{z}_{34}}{z_{12}\bar{z}_{23}}\right)^{i\lambda_{2}}\left(\frac{z_{12}\bar{z}_{13}}{z_{24}\bar{z}_{34}|z_{14}|^{2}}\right)^{\frac{i\lambda_{2}}{2}}\, \notag\\
\left(\frac{z_{12}^{1/2}\bar{z}_{12}^{1/2}z_{14}^{1/2}\bar{z}_{14}^{1/2}z_{24}^{1/2}\bar{z}_{24}^{1/2}}{z_{13}\bar{z}_{13}z_{23}\bar{z}_{23}z_{34}\bar{z}_{34}} \right)^{\frac{i\lambda_{3}}{3}} &=&r^{-\frac{i\lambda_{3}}{6}}(1-r)^{-\frac{i\lambda_{3}}{6}}\left(\frac{z_{24}\bar{z}_{14}}{z_{23}\bar{z}_{13}}\right)^{i\lambda_{3}}\left(\frac{z_{12}\bar{z}_{13}}{z_{24}\bar{z}_{34}|z_{14}|^{2}}\right)^{\frac{i\lambda_{3}}{2}}\, \notag\\
    \left( \frac{z_{12}^{1/2}\bar{z}_{12}^{1/2}z_{13}^{1/2}\bar{z}_{13}^{1/2}z_{23}^{1/2}\bar{z}_{23}^{1/2}}{z_{14}\bar{z}_{14}z_{24}\bar{z}_{24}z_{34}\bar{z}_{34}}\right)^{\frac{i\lambda_{4}}{3}} &=&r^{-\frac{i\lambda_{4}}{6}}(1-r)^{-\frac{i\lambda_{4}}{6}}\left(\frac{z_{12}\bar{z}_{13}}{z_{24}\bar{z}_{34}|z_{14}|^{2}}\right)^{\frac{i\lambda_{4}}{2}}\, 
\end{eqnarray}
and
\begin{eqnarray}
   \left( \frac{z_{12}^{5}\bar{z}_{34}^{5}}{z_{13}z_{14}z_{23}z_{24}z_{34}^{7}\bar{z}_{12}^{7}\bar{z}_{13}\bar{z}_{14}\bar{z}_{23}\bar{z}_{24}}\right)^{1/3}=\left(r^{-11}(1-r)^{-5}\right)^{\frac{1}{3}}\,.
\end{eqnarray}
Using these expressions in \eqref{expansion z}, we obtain
\begin{eqnarray}\label{z factors}
     & & \prod_{i<j}^4 z_{i j}^{\frac{h}{3}-h_i-h_j} \bar{z}_{i j}^{\frac{\bar{h}}{3}-\bar{h}_i-\bar{h}_j}=r(1-r)^{-1}r^{-\frac{14-\beta}{3}}(1-r)^{-\frac{2-\beta}{3}}\frac{\bar{z}_{34}^{5}z^{3}_{24}|z_{14}|^{2}|z_{24}|^{2}|z_{34}|^{4}}{\bar{z}_{12}z_{34}^{2}\bar{z}_{13}^{3}|z_{23}|^{6}}\notag\\
     &&\ \ \ \ \ \ \ \ \times\, \left(\frac{z_{24}\bar{z}_{34}}{z_{12}\bar{z}_{13}}\right)^{i\lambda_{1}}\left(\frac{z_{14}\bar{z}_{34}}{z_{12}\bar{z}_{32}}\right)^{i\lambda_{2}}\left(-\frac{z_{24}\bar{z}_{14}}{z_{23}\bar{z}_{13}} \right)^{i\lambda_{3}}\left(\frac{z_{12}\bar{z}_{13}}{z_{24}\bar{z}_{34}|z_{41}|^{2}}\right)^{4-\beta}\,.
\end{eqnarray}
We notice that the right hand side of the above expression, except for $r(1-r)^{-1}$, is the same factor which appears both in the celestial amplitude of gravitons in superstrings and in its field theory version in supergravity. This permits us to write the 4-graviton 1-loop celestial amplitude as proportional to
\begin{eqnarray}\label{Finally celestial amplitude}
2^{3-\beta}\kappa^{2}_{10}\,\delta(r-\bar{r})\Theta(r-1)\, r^{\frac{14-\beta}{3}}(1-r)^{\frac{2-\beta}{3}}\mathcal{J}(r, \beta)\,
\prod_{i<j}^4 z_{i j}^{\frac{h}{3}-h_i-h_j} \bar{z}_{i j}^{\frac{\bar{h}}{3}-\bar{h}_i-\bar{h}_j}
\,,
\end{eqnarray}
where $\mathcal{J}(r, \beta)$ contains the energy-integral, and it will depend on whether we are working with an amplitude in string or in field theory.

\end{document}